\documentclass[preprintnumbers,amsmath,amssymb,superscriptaddress,pra,twocolumn]{revtex4-2}

\include{amsmath}
\usepackage{graphicx}
\usepackage{subfigure}
\usepackage{amsmath}
\usepackage{amsfonts}
\usepackage{color}
\usepackage[latin1]{inputenc}

\newcommand{\therm}{\mathrm{th}}

\begin{document}

\title[]{Nonclassical photon statistics in two-tone continuously driven optomechanics}
\author{K. B{\o}rkje}
\author{F. Massel}
\affiliation{Department of Science and Industry Systems, University of
  South-Eastern Norway, PO Box 235, NO-3603 Kongsberg, Norway}
\author{J.~G.~E. Harris}
\affiliation{Department of Physics, Yale University, 217 Prospect Street, New Haven, Connecticut 06520, USA}
\affiliation{Department of Applied Physics, Yale University, 15 Prospect Street, New Haven, Connecticut 06520, USA}
\date{\today}

\begin{abstract}
In cavity optomechanics, single photon detection of Raman scattered photons can be a useful tool for observing nonclassical features of both radiation and motion. While this tool has been employed in experiments with pulsed laser driving of a cavity mode, it has not been readily accessible to experiments with continuous and constant laser driving. To address this, we present a study of a standard optomechanical system where the cavity mode is continuously driven at two different frequencies, and where sideband photons are detected by single photon detectors after frequency filtering the output from the cavity mode around its resonance frequency. We first derive the normalized second order coherence associated with the detected photons, and show that it contains signatures of the quantum nature of the mechanical mode which would be absent with only single-tone driving. To identify model-independent nonclassical features, we derive two inequalities for the sideband photon statistics that should be valid in any classical model of the system. We show that these inequalities are violated in the proposed setup. This is provided that the average phonon occupation number of the mechanical mode is sufficiently small, which in principle can be achieved through sideband cooling intrinsic to the setup. Violation of the first inequality means that there is no well-defined probability distribution of the Glauber-Sudarshan type for the cavity mode. In contrast, a violation of the second inequality means that there is no joint probability distribution for the cavity mode at two times separated by a finite interval, which originates from the noncommutativity of the motional quadratures of the mechanical mode. The proposed setup thus employs a mechanical oscillator in order to generate a steady-state source of nonclassical radiation.

%
\end{abstract}

\maketitle

\twocolumngrid

\section{Introduction}

The coupling of macroscopic mechanical resonators to optical or microwave cavity fields has become a useful experimental platform for testing quantum mechanics of composite degrees of freedom with large masses. The minute radiation pressure interaction can be amplified by coherently driving a cavity mode at sufficiently large power, leading to an effectively linear interdependence between motional degrees of freedom and the cavity mode's field fluctuations. The dynamics of such systems can only generate Gaussian states, which severely limits the possibility of distinguishing quantum and classical behaviour. Nevertheless, strong experimental evidence of the quantum nature of various macroscopic mechanical systems has been produced in the past decade, including demonstrations of zero point motion \cite{Brahms2012PRL,Safavi-Naeini2012PRL,Weinstein2014PRX,Lecocq2015NatPhys,Purdy2015PRA,Underwood2015PRA} and quantum entanglement \cite{Palomaki2013Science,Ockeloen-korppi2018Nature,Mercier2021Science,Kotler2021Science}.

A useful tool in going beyond linear dynamics in cavity optomechanics is to take advantage of projective measurements. Detecting individual photons in the mechanically induced sidebands of the coherent drive, {\it i.e.}, so-called Stokes/anti-Stokes photons or Raman photons, can give access to non-Gaussian states due to the measurement's backaction on the system. This technique requires frequency filtering of the cavity output in order to remove the large amount of photons at the carrier frequency. This has been achieved with mechanical modes of microresonators having resonance frequencies in the GHz regime \cite{Cohen2015Nature,Riedinger2016Nature,Enzian2021PRL}, with an acoustic mode of helium with frequency around 300 MHz \cite{Yu2021}, and lately, even with flexural dielectric membrane modes in the MHz regime \cite{Galinskiy2020}. 

Detection of individual sideband photons has been employed to demonstrate nonclassical phonon statistics with photonic crystal nanobeams \cite{Riedinger2016Nature,Hong2017Science},  where pulsed coherent driving at two different frequencies was used in order to sequentially detect both up- and down-converted sideband photons at the same optical detection frequency. The same technique has also been used to generate and verify entanglement between motional modes of remote mechanical nanobeams \cite{Riedinger2018Nature}, similarly to earlier experiments on optical phonons in diamond \cite{Lee2011Science}.

In the simplest case of continuous constant driving of the cavity mode with a single drive tone, the sideband photon statistics of the upper and lower sidebands measured separately are those of thermal radiation \cite{Borkje2011PRL}. Furthermore, the normalized coherences have no dependence on the average phonon occupation number of the mechanical mode. Thus, with only one detection frequency, the photon statistics do not reveal any nonclassical features. However, if both sideband frequencies can be accessed individually and their cross-correlation can be measured, a violation of classical statistics can in principle be observed also with continuous driving \cite{Borkje2011PRL}.

In this article, we consider a standard optomechanical system where the cavity mode is continuously and coherently driven at two separate frequencies, one red- and one blue-detuned from the cavity resonance by the mechanical resonance frequency. Both drives will produce sideband photons close to cavity resonance, which we assume can be detected by filtering the output of the cavity mode around its resonance frequency. 

We first show that the normalized second order coherence contains features that can be traced back to the quantum nature of the mechanical mode and its average phonon occupation number $n_m$. Next, we investigate if the system can display genuine measures of nonclassicality which cannot be explained by inaccuracies or insufficiencies in our model of the optomechanical system. We show that the observable photon statistics can indeed violate two separate classical inequalities, involving both second and third order coherences, for sufficiently small average phonon occupancies $n_m$ and for particular choices of drive strength ratios.

The violation of the first inequality we study signifies that there can be no well-defined probability distribution of the Glauber-Sudarshan type that describes the state of the (displaced) cavity mode \cite{Titulaer1965PR}. We will show that this violation can be interpreted as antibunching conditioned on a detected photon, which can occur as the sideband photons have a tendency to be emitted in well-separated pairs in the low temperature regime. This is reminiscent of emission of multiphoton (or multiphonon) bundles in cavity quantum electrodynamics \cite{SanchezMunoz2014NatPhot,Bin2020PRL,Bin2021PRL}, but differs in that the emission of pairs in the system we study  is not reliant on a system anharmonicity.

In cases where the system violates the second inequality, which can be derived from the generalized nonclassicality criterion in Ref.~\cite{Vogel2008PRL}, one may conclude that there is no well-defined {\it joint} probability distribution for the state of the cavity mode at two different times. We will argue that this can be traced back to the noncommutativity of the motional quadratures of the mechanical mode, or equivalently, that measurement of one motional quadrature will always disturb the orthogonal quadrature according to quantum mechanics. 

This study thus provides a technically simpler method for observing nonclassicality in optomechanical systems compared to previous schemes. While pulsed driving has so far been a necessity in experiments on picogram mechanical objects due to absorption heating \cite{Riedinger2016Nature,Hong2017Science,Riedinger2018Nature}, continuous driving can be possible with more massive devices, such as confined volumes of helium \cite{Yu2021} or dielectric membranes \cite{Galinskiy2020}. In addition, the setup we propose generates a steady-state source of nonclassicality, which could potentially serve as a resource in quantum-enhanced sensing schemes.

This article is organized as follows. In Section \ref{sec:Setup}, we introduce the proposed experimental setup and define the model used to describe it. Next, in Section \ref{sec:PhotStat} we study the filtered sideband photon statistics resulting from this setup when assuming a thermal mechanical state. In Section \ref{sec:ModelIndep}, we present classical inequalities for photon statistics measurements and investigate under which circumstances these inequalities are violated. The assumption of a thermal state is finally justified by the analysis of the dynamics of the mechanical oscillator in Section \ref{sec:DynBack}, where we also discuss how sideband cooling intrinsic to the proposed setup can help reach the regime where the classical inequalities are violated. We conclude in Section \ref{sec:Conclusion}.

\section{Setup and model} 
\label{sec:Setup}

We consider a standard optomechanical system in which the resonance frequency of an optical cavity mode depends linearly on the displacement of a mechanical mode. This interaction is described by the radiation pressure interaction $\hat{H}_\mathrm{int} = \hbar g_0 \hat{x} \hat{a}^\dagger \hat{a}$, where $\hat{a}$ is the photon annihilation operator, $\hat{x}$ is the mechanical displacement operator in units of its zero point motion, and $g_0$ is the shift in the cavity mode's angular resonance frequency caused by a displacement equal to the zero point motion. 

The cavity mode has an angular resonance frequency $\omega_c$ and is driven by two lasers at frequencies $\omega_r = \omega_c + \Delta_c - (\tilde{\omega}_m - \delta)$ and $\omega_b = \omega_c + \Delta_c + (\tilde{\omega}_m - \delta)$. Here, $\tilde{\omega}_m$ is the effective mechanical resonance frequency, to be defined below. We note that the two drives are centered around the frequency $\omega_\mathrm{av} = \omega_c + \Delta_c$, and red- or blue-detuned from this frequency by $\tilde{\omega}_m - \delta$. We will assume $\delta > 0$ from now on, but we note that its sign is not of importance. The optomechanical interaction will lead to Raman scattering, creating sidebands at frequencies $\pm \tilde{\omega}_m$ away from the two drive tones. 

\begin{figure}[htb]
\includegraphics[width=.99\columnwidth]{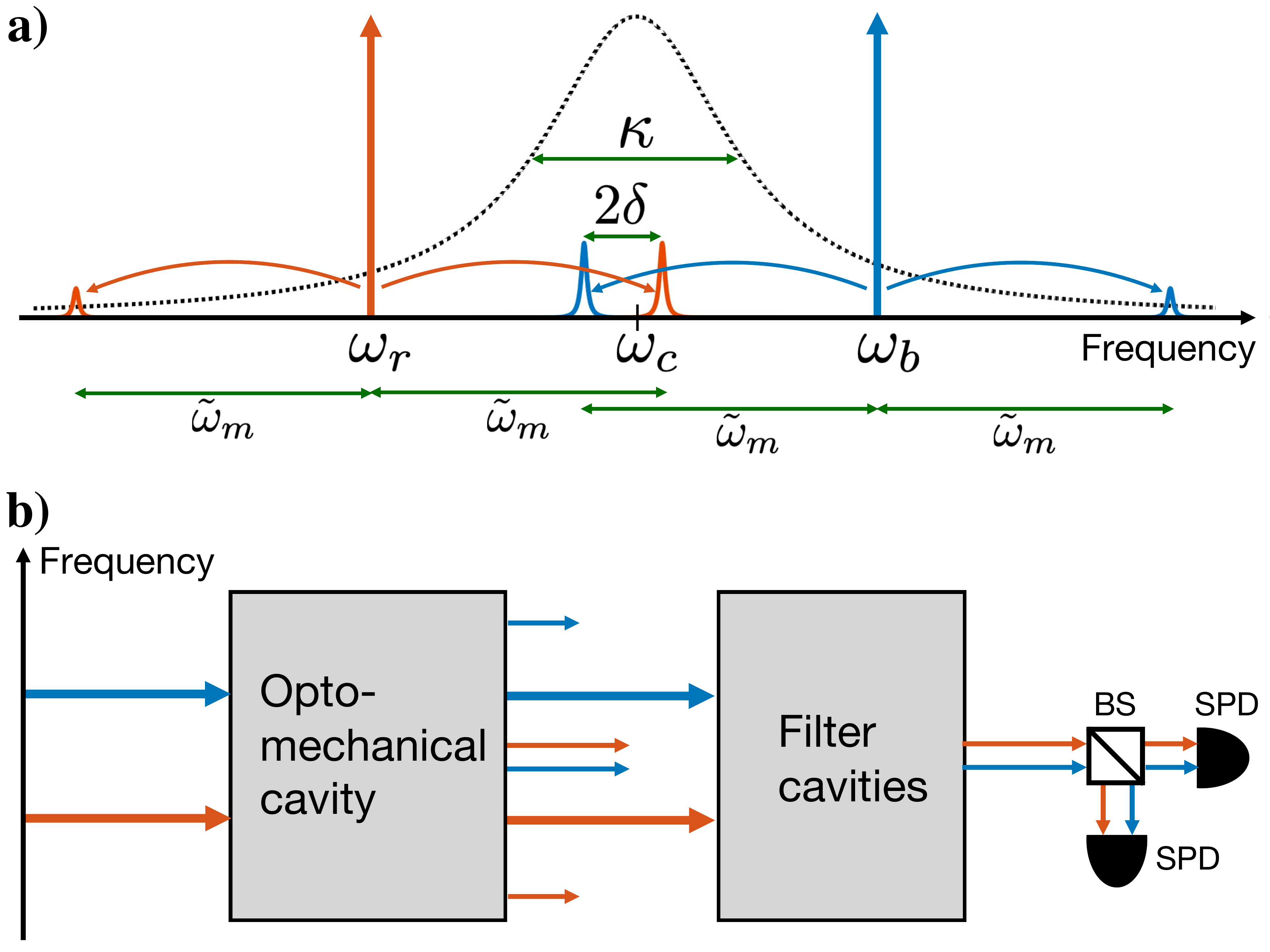}
\caption{{\bf a)} Overview of the frequencies in the proposed setup. The black (dashed) Lorentzian is the cavity response with linewidth $\kappa$. The two drive frequencies are shown by arrows, and the four mechanical sidebands are indicated as narrow Lorentzians (of width $\tilde{\gamma}$) displaced by $\pm \tilde{\omega}_m$ from the drive frequencies. {\bf b)} Schematic overview of the measurement setup. The output from the optomechanical system is sent through a set of filter cavities, in order to access only the two sidebands close to the cavity resonance frequency. After passing through a beam splitter (BS), the photon statistics of the filtered output is then measured with single photon detectors (SPDs).}
\label{fig:Setup}
\end{figure}

We consider the situation $|\Delta_c|, \delta \ll \kappa, \tilde{\omega}_m$, where $\kappa$ is the cavity energy decay rate, such that the upper sideband of the red-detuned drive and the lower sideband of the blue-detuned drive fall well within the cavity linewidth and close to the cavity resonance frequency. At the same time, we will assume that the effective mechanical linewidth $\tilde{\gamma}$ fulfils $\tilde{\gamma} \ll \delta$, such that these two sidebands are well separated by a frequency $2\delta$. This is illustrated in Figure \ref{fig:Setup}a). We note that the parameter hierarchy we propose is suitable for a variety of experimental realizations of cavity optomechanics, since the cavity decay rate $\kappa$ typically exceeds the intrinsic mechanical decay rate $\gamma$ by several orders of magnitude.

The precise value of the sideband splitting $2 \delta$ will not be important for the experiment we propose. This setup is thus different from and experimentally simpler than the special case $\delta = 0$ in which the two sidebands overlap and interfere. The latter has been considered and implemented in the contexts of back-action free quadrature measurements \cite{Thorne1978PRL,Clerk2008NJP,Hertzberg2010NatPhys,Suh2014Science,Shomroni2019NatComm}, dissipative mechanical squeezing \cite{Kronwald2013PRA_2,Wollman2015Science,Pirkkalainen2015PRL,Lecocq2015PRX}, and two-tone optomechanical instabilites \cite{Shomroni2019PRX}. We comment on this special case in Appendix \ref{sec:delta0}.

We go to a frame rotating at the average of the two drive frequencies $\omega_\mathrm{av}$, and to a frame rotating at the effective mechanical resonance frequency $\tilde{\omega}_m = \omega_m + \Delta_m$ for the mechanical mode. Here, we define $\Delta_m$ as the difference between the effective and the bare mechanical resonance frequency $\omega_m$. In terms of annihilation operators $\hat{a}$ ($\hat{b}$) for photons (phonons), the Hamiltonian then becomes
\begin{align}
H(t) & = -\hbar \Delta_c \hat{a}^\dagger \hat{a} -\hbar \Delta_m \hat{b}^\dagger \hat{b}    \\
& + \hbar  e^{i \delta t} \left(G_r \hat{a}  + G_b \hat{a}^\dagger \right) \left(e^{-2i\tilde{\omega}_m t} \hat{b} +  \hat{b}^\dagger \right) \notag \\ 
& + \hbar  e^{-i \delta t} \left(G_r \hat{a}^\dagger  + G_b \hat{a} \right) \left( \hat{b} +  e^{2i\tilde{\omega}_m t} \hat{b}^\dagger \right) \notag
\end{align}
where the coherent driving has been taken into account by displacing the cavity mode operator $\hat{a}$ at the two drive frequencies. The coupling rates $G_r$ and $G_b$ are proportional to the square root of the powers of the red and blue drive, respectively, and we can assume them real and positive without loss of generality. We have also neglected the intrinsic nonlinearity of the optomechanical interaction, assuming the experimentally relevant limit $g_0 \ll \kappa$. Finally, we have ignored any possible effect on the mechanical mode from the intensity beat note at $2 (\tilde{\omega}_m-\delta)$, since this frequency is far from any multiple of the mechanical resonance frequency $\tilde{\omega}_m$. In particular, this setup with $\delta \neq 0$ conveniently avoids potential parametric instabilities due to an intensity beat note at $2 \tilde{\omega}_m$, which has been encountered in similar experiments with $\delta = 0$ \cite{Hertzberg2010NatPhys,Steinke2013PRA}.

In order to include dissipation, we use input-output theory to find quantum Langevin equations for the annihilation operators $\hat{a}$ and $\hat{b}$ in the standard way \cite{Gardiner1985PRA}. In the adiabatic limit $\tilde{\gamma} \ll  \kappa$, we may write an implicit solution for the photon annihilation operator as
 \begin{align}
 \label{eq:aSolution}
\hat{a}(t) & =\hat{\zeta}(t) + \hat{a}_i(t) +  \hat{a}_o(t)   
\end{align}
where $\hat{\zeta}$ represents the Gaussian cavity vacuum noise due to coupling to a bath, obeying
\begin{align}
\label{eq:zetaProp}
\langle \hat{\zeta}(t) \hat{\zeta}^\dagger(t') \rangle & = e^{- \kappa |t-t'|/2 + i\Delta_c(t-t')}  
\end{align} 
and $\langle \hat{\zeta}^\dagger(t) \hat{\zeta}(t') \rangle = \langle \hat{\zeta}(t) \hat{\zeta}(t') \rangle = 0$ in the Markov approximation. We have also assumed $\hbar \omega_c \gg k_B T$, {\it i.e.}, we neglect thermal occupation of the environmental modes coupling to the cavity mode. The second term in Equation \eqref{eq:aSolution} describes the upper sideband from the red-detuned drive and the lower sideband from the blue-detuned drive, {\it i.e.}, the ``innermost'' sidebands close to cavity resonance (see Figures \ref{fig:Setup}a, \ref{fig:VirtualPhonon}a, and \ref{fig:VirtualPhonon}b), 
\begin{align}
\label{eq:aInner}
\hat{a}_i(t) = - i e^{-i \delta t} G_r \chi_c(\delta) \hat{b}(t) - i e^{i \delta t} G_b \chi_c(-\delta)  \hat{b}^\dagger(t) , 
\end{align} 
where we have defined the cavity susceptibility
\begin{align}
 \chi_c(\omega) = \frac{1}{\kappa/2 - i(\omega + \Delta_c)} .
\end{align} 
Finally, when defining $\Omega = 2\tilde{\omega}_m- \delta$, we have
\begin{align}
\hat{a}_o(t) = - i e^{-i\Omega t} G_b \chi_c(\Omega) \hat{b}(t) - i e^{i\Omega t} G_r \chi_c(-\Omega) \hat{b}^\dagger(t)  ,
\end{align} 
which describes the ``outermost'' sidebands, {\it i.e.}, the lower sideband of the red-detuned drive and the upper sideband of the blue-detuned drive - see Figure \ref{fig:Setup}a). 

In Section \ref{sec:DynBack}, we will discuss the dynamics of the mechanical mode and argue that, to a very good approximation, it is in a thermal steady state. The state is characterized by an average phonon occupation number $n_m$, and we will denote the effective mechanical energy decay rate $\tilde{\gamma}$. For the time being, we treat these as independent parameters, and show later how they depend on the various parameters of our model. This means that all correlation functions describing the mechanical mode can be expressed in terms of the second order correlation functions 
\begin{align}
\label{eq:bCorrDef}
\langle \hat{b}^\dagger(t+\tau) \hat{b}(t) \rangle & = n_m e^{-\tilde{\gamma} \tau/2} \\
\langle \hat{b}(t+\tau) \hat{b}^\dagger(t) \rangle & = \left(n_m + 1\right) e^{-\tilde{\gamma} \tau/2} , \label{eq:bCorrDef2}
\end{align}
where $\tau \geq 0$. The additional ``+1'' in the second line originates from the boson commutation relations, indicating the quantum nature of the mechanical oscillator.

\section{Second order coherence of sideband photons} 
\label{sec:PhotStat}

We will now consider the photon statistics of the two center (or ``innermost'') sidebands combined. In practice, this can be measured by frequency filtering the cavity output around the cavity resonance frequency with a filter bandwidth $B$ satisfying $\delta \ll B \ll \tilde{\omega}_m$, before the sidebands are sent to single-photon detectors. This is schematically illustrated in Figure \ref{fig:Setup}b).

A central assumption in the following will be that the photodetectors destroy all information about the frequency of a detected photon. This means that as long as the mechanical mode is not interrogated, there is no way of knowing whether a detected photon was down-converted from the blue-detuned drive or up-converted from the red-detuned drive, {\it i.e.}, a quantum superposition of a phonon creation and a phonon annihilation will occur. 

We note that such interference between up- and down-converted photons is what leads to a squeezed mechanical state in the case of $\delta = 0$ \cite{Kronwald2013PRA_2}. For $\delta \neq 0$, however, the {\it average} mechanical state stays thermal, since the squeezing angle rotates with frequency $\delta$ such that the effect of this interference averages out. This justifies why we have assumed a mechanical steady state that is invariant under time translation.
 
We start by considering the normalized second order coherence for the filtered cavity mode, which we can express as
\begin{align}
\label{eq:g2Def}
g^{(2)}(t,t+\tau) & =  \frac{\langle \hat{a}^\dagger_f(t) \hat{a}^\dagger_f(t+\tau) \hat{a}_f(t+\tau)  \hat{a}_f(t) \rangle  }{\langle \hat{a}_f^\dagger(t)   \hat{a}_f(t) \rangle \langle  \hat{a}_f^\dagger(t+\tau) \hat{a}_f(t+\tau)  \rangle } 
\end{align}
when defining 
\begin{align}
\label{eq:afDef}
\hat{a}_f(t) & = \hat{\zeta}_f(t) + \hat{a}_i(t)
\end{align}
with $\hat{\zeta}_f(t)$ the filtered cavity vacuum noise. The latter is defined in Appendix \ref{app:Filtering}, where further details on the filtering can be found.

For simplicity, we will now consider the limits $\delta/\kappa, |\Delta_c|/\kappa \rightarrow 0$, which means that we will be ignoring that the cavity susceptibility is slightly different for the two ``innermost'' sidebands. We also take the limit $\tilde{\gamma}/\delta \rightarrow 0$, which neglects any overlap between the two sidebands. We will study corrections to our results beyond these limits in Appendix \ref{app:Corrections}. The corrections turn out to be of first order in $\tilde{\gamma}/\delta$, but only of second order in $\delta/\kappa, |\Delta_c|/\kappa$. 

The optomechanical interaction results in the cavity vacuum noise $\hat{\zeta}$ becoming correlated with the mechanical mode, {\it i.e.} $\langle \hat{\zeta}(t+\tau) \hat{b}^{(\dagger)}(t) \rangle \neq 0$. This means there are nonzero terms in the numerator of $g^{(2)}(t,t+\tau)$ where $\hat{\zeta}_f(t+\tau)$ directly enters. These terms represent off-resonant virtual phonon processes where two photons, one up-shifted and one down-shifted, are created simultaneously and emitted at the frequency $\omega_\mathrm{av}$ within a time interval $\sim 1/\kappa$, as illustrated in Figure \ref{fig:VirtualPhonon}c. However, it turns out (see Appendix \ref{app:Corrections}) that these terms only give corrections of order $\delta^2/\kappa^2$ or $\delta |\Delta_c|/ \kappa^2$ to the result one finds by replacing $\hat{a}_f$ with $\hat{a}_i$ in Equation \eqref{eq:g2Def}. Thus, we will neglect these terms in the limit we consider now.
\begin{figure}[htb]
\includegraphics[width=.99\columnwidth]{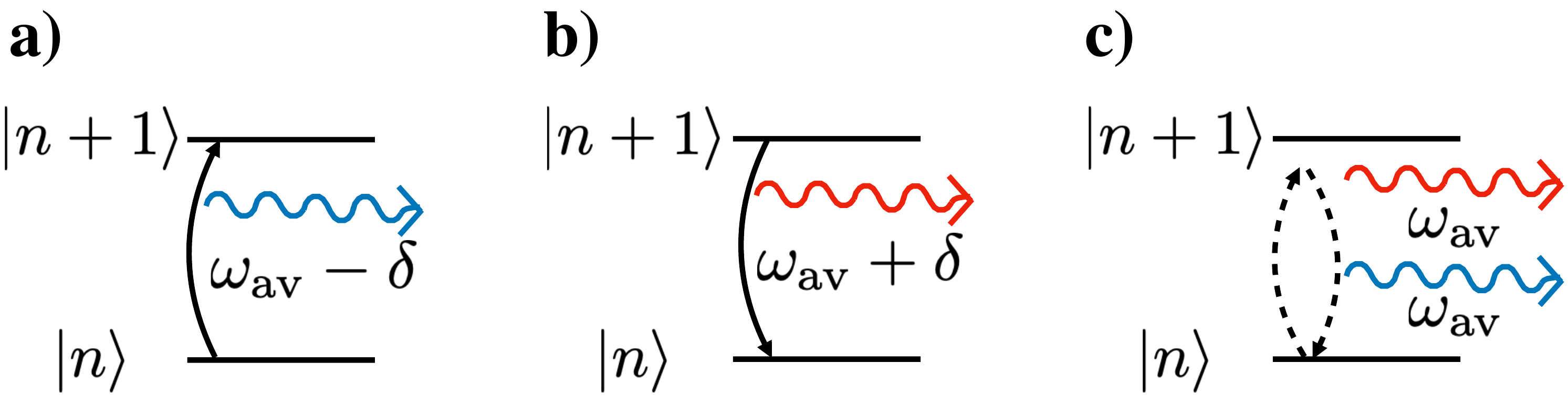}
\caption{Illustration of processes that emit sideband photons close to the cavity resonance frequency. {\bf a)} A phonon is created by emission of a down-converted photon from the blue-detuned drive at frequency $\omega_\mathrm{av} - \delta$. {\bf b)} A phonon is annihilated by emission of an up-converted photon from the red-detuned drive at frequency $\omega_\mathrm{av} + \delta$. {\bf c)} A virtual phonon is created and immediately annihilated resulting in the emission of two photons, one up-converted and one down-converted, at the same frequency $\omega_\mathrm{av}$. A process with the opposite order of phonon creation and annihilation is also possible.}
\label{fig:VirtualPhonon}
\end{figure}

Given these simplifications, we may now write down an expression for the normalized second order coherence. Recognizing that it is independent of absolute time $t$ for a thermal mechanical state and thus simplifying the notation by $g^{(2)}(t,t+\tau)  \rightarrow g^{(2)}(\tau)$, we find 
\begin{align}
\label{eq:g2Sol}
& g^{(2)}(\tau)  =  1  \\
&  \quad + e^{-\tilde{\gamma} \tau} \left(1 +  \frac{4 \beta  \big[1/4 + n_m (n_m+1) \cos (2\delta \tau)\big]  }{\big[n_m + \beta (n_m + 1)\big]^2} \right)  \notag
\end{align}
when we define the squared ratio between the optomechanical coupling constants as
\begin{align}
\label{eq:beta0Def}
\beta  & =    \left(\frac{G_b}{G_r}\right)^2 .
\end{align}
We note that the expression \eqref{eq:g2Sol} is not well-defined if both $n_m$ and $\beta$ are zero. This is reasonable since if that were the case, there would be no sideband photons to detect. A phonon occupation number $n_m$ that is strictly zero is also unphysical when taking the off-resonant sidebands into account, as will be evident in Section \ref{sec:DynBack}.

Let us first note that if we consider the case of only a single drive tone, {\it i.e.}, $G_b = 0$ ($\beta = 0$) or $G_r = 0$ ($\beta \rightarrow \infty$), we get $g^{(2)}(\tau) = 1 + e^{-\tilde{\gamma} \tau}$. This is characteristic of (classical) thermal radiation, and there is no dependence on the phonon occupation number $n_m$  \cite{Borkje2011PRL}. 

For other values of the ratio $\beta$, however, the function $g^{(2)}(\tau)$ oscillates with a period of $\pi/\delta$ and with a time-decaying amplitude $e^{-\tilde{\gamma} \tau} A$ with initial size
\begin{align}
\label{eq:ADef}
A  & =    \frac{4 \beta  n_m (n_m+1)   }{\big[n_m + \beta (n_m + 1)\big]^2} . 
\end{align}

\begin{figure}[htb]
\includegraphics[width=.99\columnwidth]{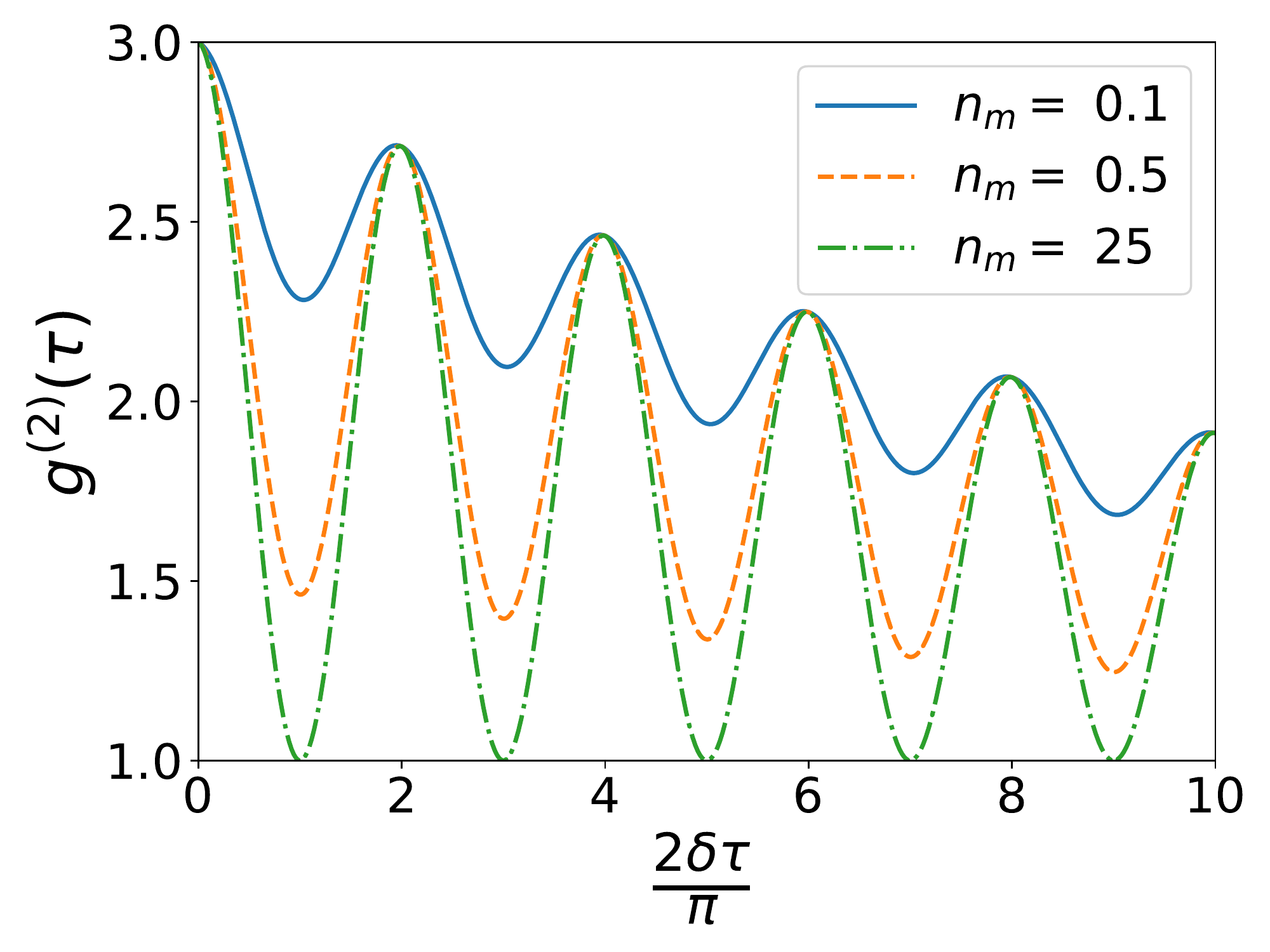}
\caption{Normalized second order coherence $g^{(2)}(\tau)$ as a function of time delay $\tau$ in the case of $\beta = 1$ and with $\tilde{\gamma}/\delta = 0.05$.}
\label{fig:g2nmPlot}
\end{figure}

\begin{figure}[htb]
\includegraphics[width=.99\columnwidth]{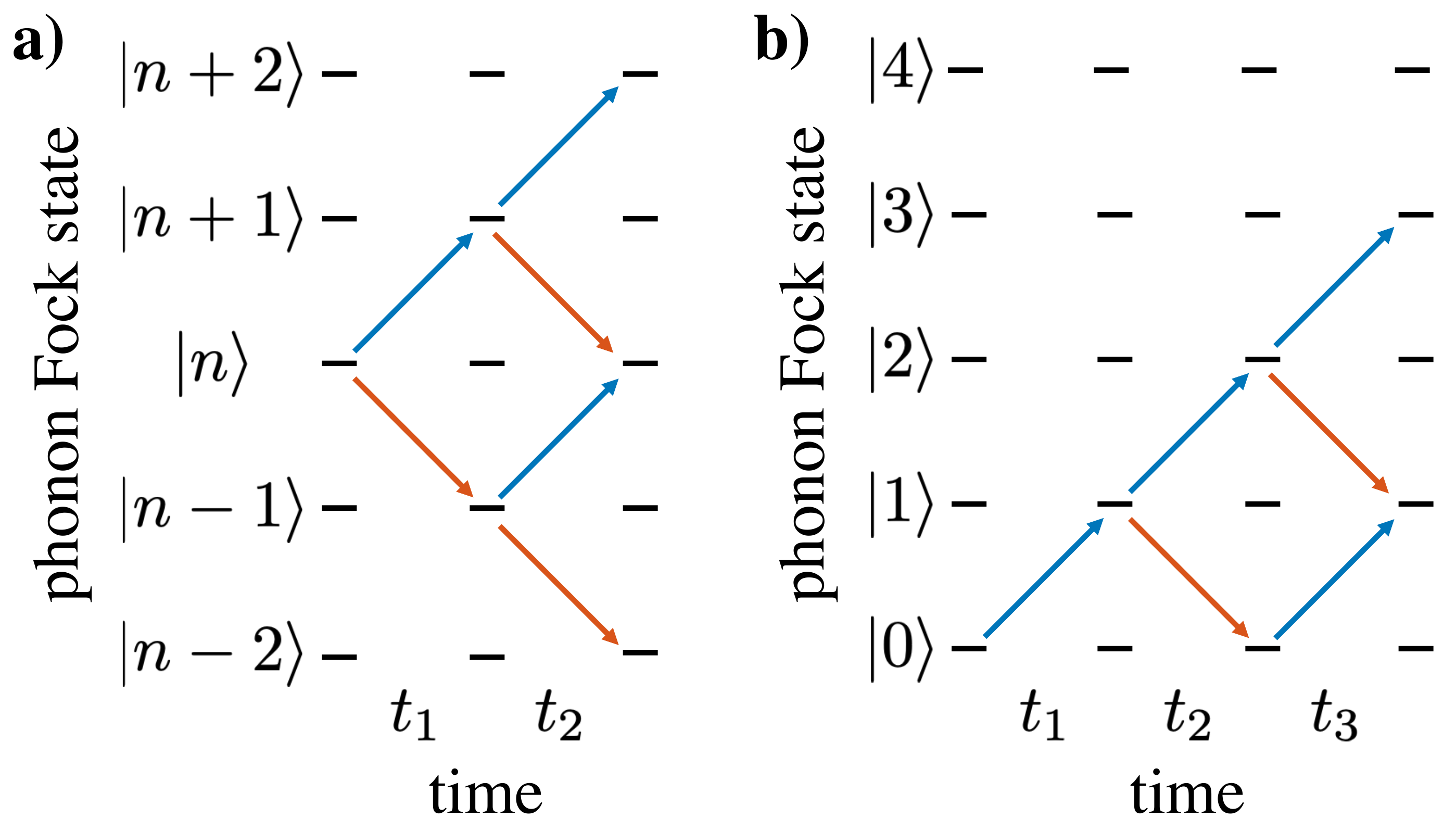}
\caption{Illustration of phonon creation and annihilation processes associated, respectively, with emission of down-converted photons from the blue-detuned drive and up-converted photons from the red-detuned drive at times $t_1$, $t_2$, ($t_3$). {\bf a)} In general, there can be interference between two paths in the Fock state ladder, corresponding to different time orderings of one up-converted and one down-converted photon. Here, we imagine starting from the $n$'th Fock state. In reality, the oscillator is initially in a thermal state, {\it i.e.}, a mixed state of different phonon numbers. {\bf b)} When starting from the phonon ground state, there are interfering paths in the phonon Fock state ladder for three detected sideband photons, but not for only two detected photons.}
\label{fig:Levels}
\end{figure}

In Figure \ref{fig:g2nmPlot}, we plot the normalized second order coherence function in Equation \eqref{eq:g2Sol} for the special case of $\beta = 1$, {\it i.e.}, equal strengths for the two drives, and for different values of the phonon occupation number $n_m$.
In the classical limit of large $n_m$, we can interpret these oscillations as interference between classically correlated sidebands. In a quantum interpretation, we can think of the oscillations as interference between a process where a phonon is first created and subsequently destroyed and the opposite process. This is illustrated in Figure \ref{fig:Levels}a). We also note that the oscillations disappear in the limit $n_m \rightarrow 0$. The oscillator is then most likely in the ground state before the first photon is detected, which means there are no two-step paths in the phonon Fock state ladder that can interfere.

\begin{figure}[htb]
\includegraphics[width=.99\columnwidth]{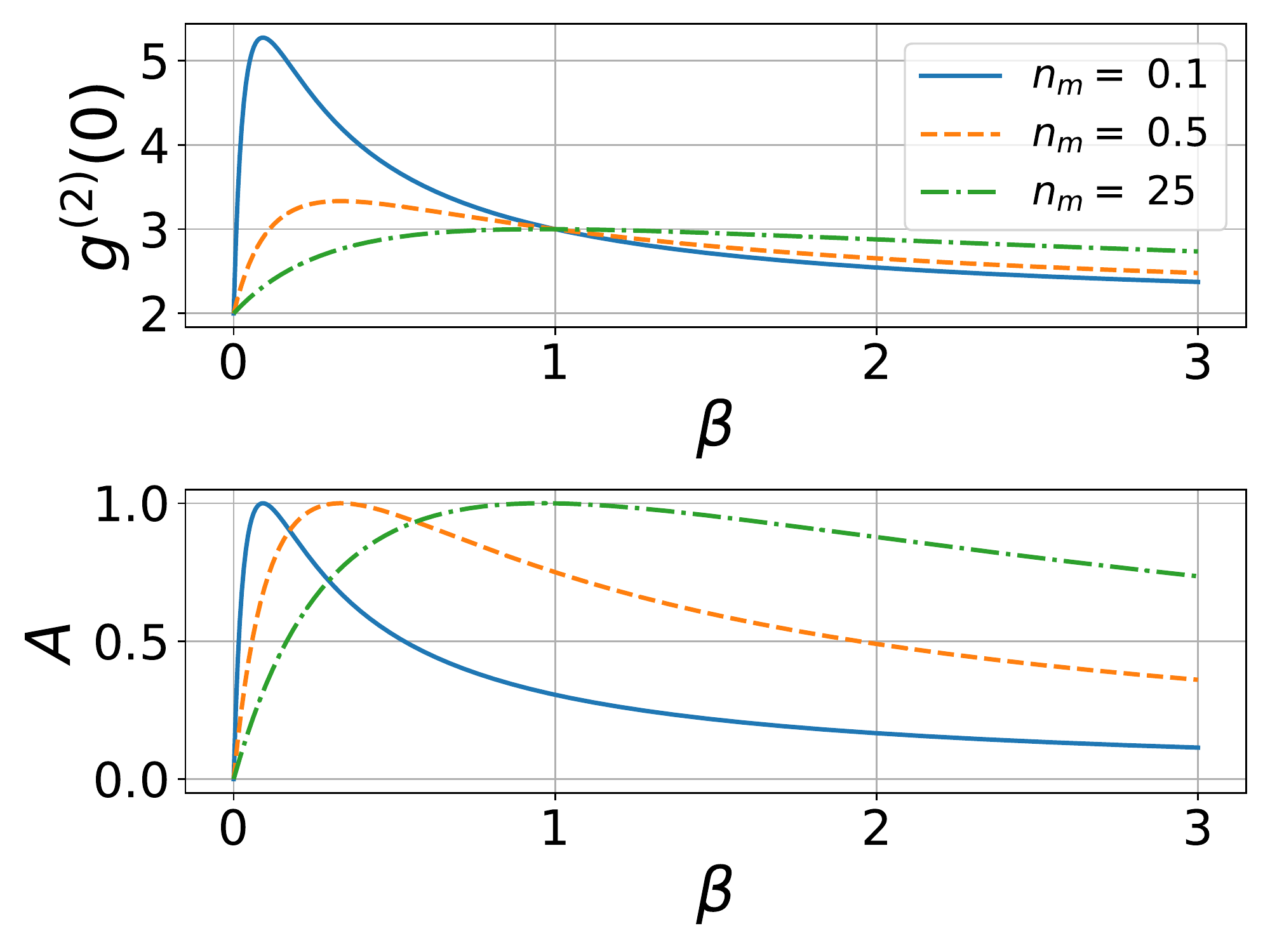}
\caption{{\it Upper panel}: Normalized second order coherence $g^{(2)}(0)$ at zero time delay as a function of $\beta$. {\it Lower panel}: Initial oscillation amplitude $A$ as a function of $\beta$.}
\label{fig:AmpbetaPlot}
\end{figure}

To explore a wider range of the drive strength ratio $\beta$, we plot the normalized second order coherence function in Equation \eqref{eq:g2Sol} at zero time delay $g^{(2)}(0)$ and the amplitude $A$ as a function of the ratio $\beta$ for different values of the phonon occupation number $n_m$ in Figure \ref{fig:AmpbetaPlot}.
We observe that $g^{(2)}(0) = 2$ for $\beta = 0$ and $g^{(2)}(0)= 3$ for $\beta = 1$ irrespective of the value of $n_m$. The former is consistent with thermal radiation, as mentioned above. The latter can be understood by realising that for $\beta = 1$, the filtered cavity mode only couples to a single quadrature of the mechanical oscillator \cite{Thorne1978PRL,Clerk2008NJP} at a time, {\it i.e.}, we can then write $\hat{a}_i(t) \propto \hat{X}_{\delta t} (t) $, where
\begin{align}
\label{eq:Quad}
\hat{X}_{\delta t} (t)  = \frac{1}{\sqrt{2}} \left( e^{- i \delta (t - t_0) } \hat{b}(t) + e^{ i \delta (t - t_0) } \hat{b}^\dagger(t) \right) 
\end{align}
and $\delta t_0$ is the complex phase of $\sqrt{\chi^\ast_c(-\delta)\chi_c(\delta)}$. This gives $g^{(2)}(0)= \langle \hat{X}_{0}^4(0) \rangle/\langle \hat{X}_{0}^2(0) \rangle^2 = (4 - 1)!! = 3$ for any Gaussian quadrature probability distribution by using Isserlis' theorem.

Figure \ref{fig:AmpbetaPlot} also shows that $g^{(2)}(0)$ can become very large in the limits $n_m \ll 1$ and $\beta \ll 1$, i.e., when the probability of the mechanical oscillator being in the ground state is close to unity. A detected photon is then most likely a down-converted photon from the blue-detuned drive that excites the oscillator from the ground state to the first excited state. These processes happen at a rate $\sim G_b^2/\kappa$, although the rate of {\it detected} photons will of course also depend on the detection efficiency. Conditioning on one such photon detection gives a large increase in the probability of an immediate second photon detection and thus a large $g^{(2)}(0)$ \cite{Borkje2011PRL}. The reason is that an up-converted photon from the red-detuned drive can then return the oscillator to the ground state, which is more likely than further exciting the oscillator since $G_r \gg G_b$. 

In other words, in the limits $n_m \ll 1$ and $\beta \ll 1$, the detected photons tend to come in well-separated pairs, with a pair consisting of one down-converted followed by one up-converted photon. More precisely, the ratio between the time scale between two photons in a pair and the time scale between two pairs is $\beta \ll 1$. We note that despite this {\it tendency} of well-separated pairs of photons, not all photons necessarily come in pairs since the oscillator can both be excited and deexcited through its coupling to other degrees of freedom, {\it i.e.}, its environment. However, the normalization of $g^{(2)}$ ensures that it nevertheless captures this tendency.
 
Let us now return to the special case $\beta = 1$, where the filtered output field is proportional to the mechanical quadrature $\hat{X}_{\delta t}$ at time $t$. This means that at two different times $t,t'$ separated by $t' - t = \pi/2\delta$, the cavity is susceptible to orthogonal quadratures, $\hat{X}_{\delta t}$ and $\hat{X}_{\delta t + \pi/2}$. For a mechanical steady state that is Gaussian and rotationally invariant in phase space, any deviation of $g^{(2)}(\pi/2\delta)$ from unity can then be traced back to a nonzero commutator $[\hat{X}_{\delta t}(t) , \hat{X}_{\delta t+\pi/2}(t')] \approx i$ between orthogonal quadratures. The physical interpretation of this is that of quantum measurement backaction. Detection of a photon at time $t$ translates to a measurement of the mechanical oscillator along a particular direction in phase space, which disturbs the orthogonal quadrature.

We also note that the phonon occupation number $n_m$ is accessible from measurements of $g^{(2)}(\tau)$, since
\begin{align}
\label{eq:nmFromg2}
\frac{g^{(2)}(0) - 2}{1 + e^{-\pi \tilde{\gamma}/2\delta} - g^{(2)}(\pi/2\delta)} = \frac{(n_m + 1/2)^2}{(n_m + 1/2)^2 - 1/2}
\end{align}
which is independent of $\beta$. In the limit $\tilde{\gamma}/\delta \rightarrow 0$, the parameter $\delta$ can be determined from the positions along the time axis of the local minima of $g^{(2)}(\tau)$, whereas $\tilde{\gamma}$ can be determined from the decay envelopes. 

\section{Model-independent nonclassicality} 
\label{sec:ModelIndep}

Although the second order coherence $g^{(2)}(\tau)$ in Equation \eqref{eq:g2Sol} contains features that stem from the quantum nature of the mechanical oscillator, it was derived under the {\it assumptions} of a thermal mechanical state ({\it i.e.}, perfectly Gaussian and rotationally symmetric in phase space) and the absence of technical laser noise. A pertinent question is thus whether this system can produce photocurrent statistics that cannot be explained by any classical model. 

To explore genuine nonclassical features, we now define the normalized third order coherence
\begin{align}
\label{eq:g3Def}
g^{(3)}(t,t,t+\tau) & =  \frac{\langle \hat{a}_f^{\dagger \, 2}(t) \hat{a}_f^\dagger(t+\tau) \hat{a}_f(t+\tau)  \hat{a}_f^2(t) \rangle  }{\langle \hat{a}_f^\dagger(t)   \hat{a}_f(t) \rangle^2 \langle  \hat{a}_f^\dagger(t+\tau) \hat{a}_f(t+\tau)  \rangle } ,
\end{align}
where the numerator is proportional to the probability rate of detecting two photons at the same time $t$ followed by one photon at time $t +\tau$. 

As with the second order coherence, we will again simplify the calculation by replacing $\hat{a}_f$ with $\hat{a}_i$ in \eqref{eq:g3Def}. In Appendix \ref{app:Corrections}, we show that the terms neglected in general give corrections to $g^{(3)}$ of first order in $\delta/\kappa$. However, for the particular delay times $\tau$ that we will consider below, one can show that the corrections are in fact only of second order in $\delta/\kappa$.

For a thermal mechanical state, $g^{(3)}(t,t,t+\tau)$ will also feature oscillations with delay time $\tau$ of period $\pi/\delta$, and thus local mimima at odd multiples of $\pi/2\delta$, due to destructive interference. While it is straightforward to calculate the full expression for $g^{(3)}(t,t,t+\tau) \rightarrow g^{(3)}(\tau)$ (see Appendix \ref{app:CalcDetails}), we will focus on the two special cases $\tau = 0$, which gives
\begin{align}
\label{eq:g3EqualTime}
g^{(3)}(0) = 9 g^{(2)}(0) - 12 
\end{align}
for any Gaussian state, and $\tau = \pi/2\delta$, {\it i.e.}, at the first local minimum, where
\begin{align}
\label{eq:g3DelayTime}
& g^{(3)}\left(\frac{\pi}{2\delta}\right) = 6 +  \frac{\beta }{\left[n_m + \beta (n_m + 1)\right]^2} \\
& \times \left[8 - 3(2 n_m + 1)^2   +  4  (2 n_m +1 ) \frac{n_m - \beta (n_m + 1) }{n_m + \beta (n_m + 1)} \right]  \notag
\end{align}
for a thermal state and in the limit $\tilde{\gamma}/\delta \rightarrow 0$. 

We now define the quantity
\begin{align}
\label{eq:KDef}
K(t,t+\tau) & =  \frac{g^{(3)}\left(t,t,t + \tau\right) }{\left[g^{(2)}\left(t,t + \tau\right)  \right]^2} .
\end{align}
For a thermal mechanical state, $K(t,t+\tau) \rightarrow K(\tau)$ is also independent of the absolute time $t$, but we emphasize that the nonclassicality criteria presented below are valid in the general case and do not rely on any assumptions about the nature of the optomechanical system.

\begin{figure}[htb]
\includegraphics[width=.99\columnwidth]{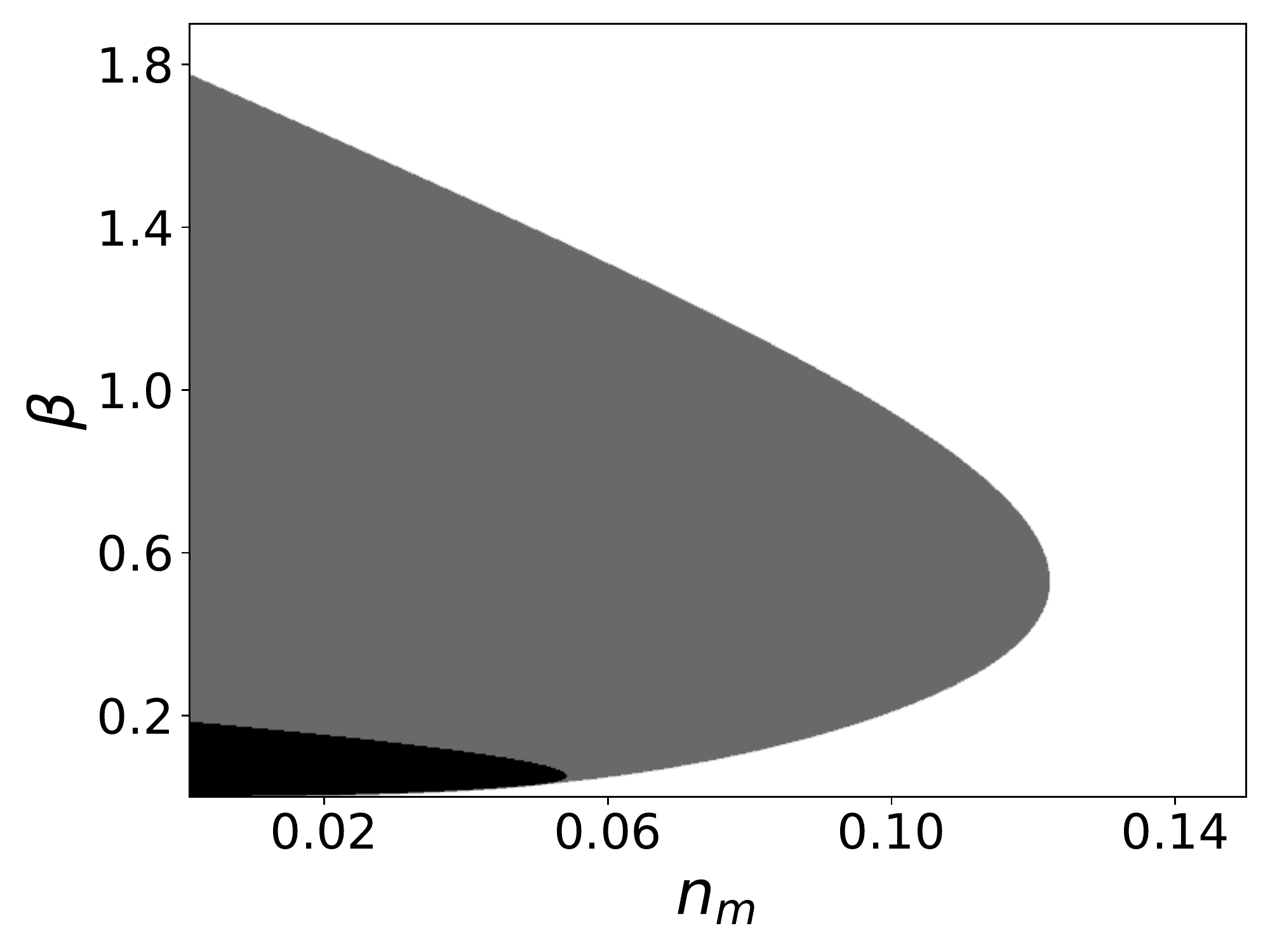}
\caption{{\it Black}: Region of parameter space where both inequalities \eqref{eq:K0Ineq} and \eqref{eq:KDelayIneq} are violated. {\it Dark gray}: Region of parameter space where the inequality \eqref{eq:KDelayIneq} is violated.}
\label{fig:KDelayPlot}
\end{figure}

In a state where the filtered cavity mode $\hat{a}_f $ can be represented by a positive-definite Glauber-Sudarshan distribution $P(\alpha)$ of the coherent complex cavity amplitude $\alpha$, one can show \cite{Titulaer1965PR} that 
\begin{align}
\label{eq:K0Ineq}
K(t,t) & \geq 1 . 
\end{align}
We can for example think of $P(\alpha)$ as describing the state of a filter cavity whose input is the output from the optomechanical cavity in a cascaded setup \cite{Gardiner1993PRL,Carmichael1993PRL} (see Appendix \ref{app:Filtering}). According to \eqref{eq:g3EqualTime}, this classicality criterion \eqref{eq:K0Ineq} is violated for a Gaussian state if $g^{(2)}(0) > (9 + \sqrt{33})/2 = 7.37$, which can occur in the system we have considered for sufficiently small $n_m$ and $\beta$. In Figure \ref{fig:KDelayPlot}, we show the parameter region (black color) where \eqref{eq:K0Ineq} is violated. We observe that it requires a phonon occupation number $n_m \lesssim 0.054$ when choosing an optimal drive ratio $\beta = 0.05$.

The violation of \eqref{eq:K0Ineq} reflects that the cavity mode is in a squeezed state with an average photon occupation number much smaller than unity. As discussed above, this means that photons tend to come in pairs (one down-converted followed by one up-converted), but that there is little overlap in time between the different pairs. Thus, conditioned on having detected one photon, the probability of immediately detecting two more photons is relatively small. In fact, we may write $K(t,t)= g_c^{(2)}(t,t)$, where the subscript $c$ indicates that it is the normalized second order coherence in the state {\it conditioned} on one photon detection. This means that we can regard the violation of \eqref{eq:K0Ineq} as conditional antibunching.

To see this in a different way, let us imagine that the mechanical mode is initially in the ground state. The first photon detection will then produce a single phonon Fock state. For $\beta \ll 1$, $\hat{a}_i(t)$ is approximately proportional to $\hat{b}(t)$, such that the filtered photon statistics is almost the same as the phonon statistics, which will feature antibunching for a single phonon Fock state.

We note that \eqref{eq:K0Ineq} cannot be violated in the special case $\beta = 1$. This is as expected, since $K(t,t)$ can in that case be expressed in terms of single-time expectation values of only one mechanical quadrature, for which there exists a well-defined Gaussian probability distribution.

For finite time delay $\tau$, we can also derive an inequality that must be satisfied by a mode that has a well-defined {\it joint} probability distribution $P(\alpha_1,\alpha_2)$ \cite{Vogel2008PRL,Krumm2017PRA} of coherent complex amplitudes $\alpha_1$ and $\alpha_2$ at times $t$ and $t + \tau$, respectively. The inequality 
\begin{align}
\label{eq:KDelayIneq}
K(t,t+\tau) & \geq 1 
\end{align}
can be derived directly from the Cauchy-Bunyakovski-Schwarz inequality or from the generalized multimode classicality criterion derived in Ref.~\cite{Vogel2008PRL}. In the system we have considered, and for time delay $\tau = \pi/2\delta$, the inequality \eqref{eq:KDelayIneq} is violated in a larger region of parameter space than the equal-time inequality \eqref{eq:K0Ineq}, as shown in Figure \ref{fig:KDelayPlot} (dark gray color). In this case, nonclassicality can be observed for $n_m \lesssim 0.12$ at an optimal $\beta = 0.53$. We also note that measurement of $K\left(\pi/2\delta\right)$ only requires two-photon coincidence detection, unlike $K(0)$, which requires three-photon coincidence measurements.

\begin{figure}[htb]
\includegraphics[width=.99\columnwidth]{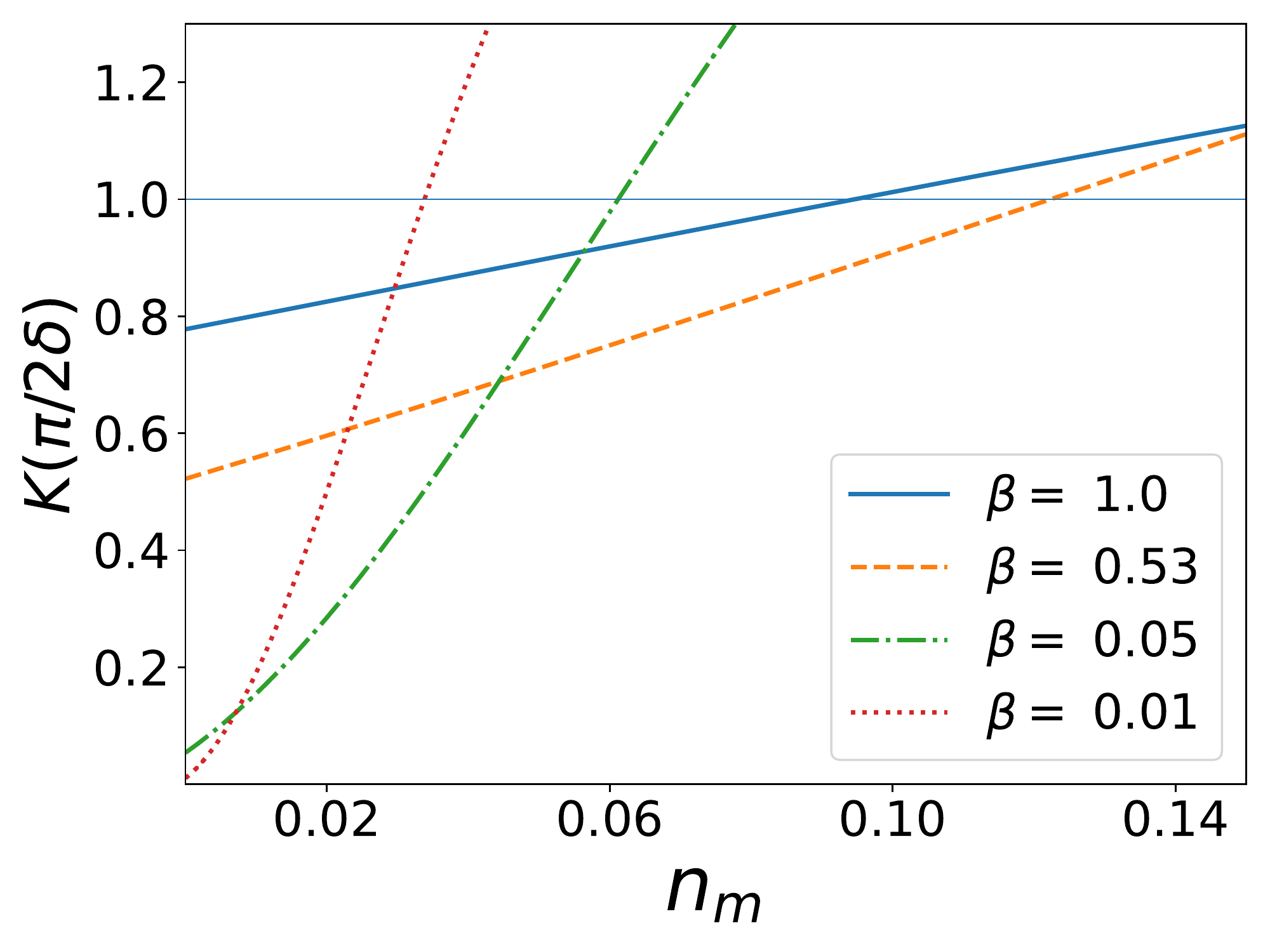}
\caption{The quantity $K(\pi/2\delta)$ as a function of average phonon occupation number $n_m$. We observe that the inequality \eqref{eq:KDelayIneq} is clearly violated for sufficiently small $n_m$. The optimal choice of drive strength ratio in order to observe violation of \eqref{eq:KDelayIneq} is $\beta = 0.53$.}
\label{fig:KvsnmPlot}
\end{figure}

In Figure \ref{fig:KvsnmPlot}, we plot $K(\pi/2\delta)$ as a function of the phonon occupation number $n_m$ for different values of the drive ratio $\beta$. We observe that the inequality \eqref{eq:KDelayIneq} is clearly violated for sufficiently small occupation numbers, which means that it should be observable if this parameter regime can be accessed. One would then be able to conclude that there can be no joint probability distribution for the cavity field for times separated by $\pi/2\delta$, even in cases where all single-time cavity expectation values can be calculated from a well-defined probability distribution, {\it e.g.}, for $\beta = 1$.

The violation of the inequality \eqref{eq:KDelayIneq} can be understood from the fact that starting from the oscillator ground state, the three photon detection amplitude can be reduced due to destructive interference for an appropriate delay time, whereas the two-photon amplitude cannot. This is illustrated in Figure \ref{fig:Levels}b. This Figure also helps motivate why we only consider two distinct detection times, and not three, in the definition of the third order coherence \eqref{eq:g3Def}. In the regime where quantum effects are significant, the mechanical oscillator is with high probability in Fock state $|1\rangle$ after the first photon detection, such that a nonzero time delay between the first and the second photon does not lead to any interference effects, only mechanical decay.

It is remarkable that the interaction with the mechanical mode can give rise to these genuinely nonclassical effects, since, when averaging over the environment and the measurement record, the mechanical mode is in a thermal steady state which can be characterized by quasiprobability distributions that are always positive. The explanation is that the {\it ordered} mechanical expectation values which appear in the nonclassicality measure $K(t,t+\tau)$ cannot be calculated from a single such distribution without invoking the quantum commutation relation between mechanical quadratures.

\section{Dynamical backaction} 
\label{sec:DynBack}

The quantum signatures we have discussed are observable in the limit of small average phonon occupation numbers $n_m$. They do not, however, depend on the absolute values of the coupling rates $G_r$ and $G_b$, only their ratio through $\beta$. One possible way to observe these features is thus to cool a high-frequency mechanical oscillator close to the quantum ground state such that $n_\mathrm{th} \ll 1$, where
\begin{align}
\label{eq:nth}
n_\therm & = \frac{1}{e^{\hbar \omega_m/(k_B T_\mathrm{eff})} - 1}  
\end{align}
is the thermal occupation number of the oscillator's effective environmental bath with temperature $T_\mathrm{eff}$. In this case, one could use small coupling rates $G_r, G_b$ such that the thermal mechanical state is essentially unperturbed, {\it i.e.} $n_m \approx n_\therm \ll 1$. We do note, however, that the absolute values of the coupling rates determine the photon flux arriving at the detector, such that there is a limit to how small they can be and still provide adequate statistics, depending on the dark current noise of the detectors.

The cooling to $n_\mathrm{th} \ll 1$ could be achieved either by conventional refrigeration, additional laser cooling with a third laser drive, or both. In fact, it would even be possible to use the same cavity mode $\hat{a}$ for cooling with a third drive tone (at a non-optimal, red-detuned frequency), provided that neither the cooling tone nor its sidebands make it through the frequency filter. 

Another possibility for reaching the regime $n_m \ll 1$ is to exploit sideband cooling intrinsic to the two-tone setup by operating at small values of $\beta$, in which case up-conversion from the red-detuned drive will cool the oscillator mode more than down-conversion from the blue-detuned drive will heat it. This requires that the system is in the resolved sideband regime $\omega_m > \kappa$. We note that several of the experimental setups where single sideband photon detection have been implemented are indeed in this regime \cite{Cohen2015Nature,Riedinger2016Nature,Enzian2021PRL,Yu2021}.

We will now take into account the mechanical oscillator dynamics in order to investigate in which parameter regime, {\it i.e.}, for which values of $G_r$, $\beta$, and $n_\therm$, the nonclassical features discussed in Section \ref{sec:ModelIndep} can be observed.  

Using the adiabatic solution \eqref{eq:aSolution} gives the following Langevin equation for the phonon annihilation operator:
\begin{align}
\label{eq:bEquation}
\dot{\hat{b}} & = -\frac{\tilde{\gamma}}{2} \hat{b} - \mu e^{2i\delta t} \hat{b}^\dagger +  \sqrt{\gamma} \hat{\eta} \\
& - i e^{i \delta t} \left(G_r \hat{\zeta}  + G_b \hat{\zeta}^\dagger \right)  -  i e^{ i \Omega t} \left(G_b \hat{\zeta} + G_r \hat{\zeta}^\dagger   \right) .\notag
\end{align}
In the limits $\delta/\kappa, |\Delta_c|/\kappa \rightarrow 0$, the effective mechanical linewidth is
\begin{align}
\tilde{\gamma} & = \gamma \left[1 + (1 - s ) (C_r - C_b)  \right]
\end{align}
where $\gamma$ is the intrinsic mechanical linewidth, we have defined
\begin{align}
\label{eq:sDef}
s & = \frac{1}{1 + (4\omega_m/\kappa)^2} ,
\end{align}
and we have introduced the optomechanical cooperativities
\begin{align}
\label{eq:CDef}
C_j & = \frac{4G_j^2}{\kappa \gamma} ,
\end{align}
with $j = r,b$. The cooperativities are measures of how strongly the mechanical and optical degrees of freedom interact relative to their intrinsic decay rates. In order for the linearized model to be valid, we must have that $\tilde{\gamma} > 0$ to avoid instability. For $C_r, C_b \ll 1$, this is always satisfied. For $C_r, C_b \gg 1$, it is always satisfied for $\beta \leq 1$. 

The off-resonant term in Equation \eqref{eq:bEquation} proportional to $\mu \sim {\cal O} (\gamma C_r \sqrt{\beta} \Delta_c/\kappa)$ can safely be neglected in the limits $\tilde{\gamma}/ \delta, |\Delta_c|/\kappa \rightarrow 0$ we consider (see also Appendix \ref{app:Mech}). 

The operator $\hat{\eta}$ describes thermal and quantum noise from the mechanical mode's effective bath, which we assume to be Gaussian and where
\begin{align}
\label{eq:etaProp}
\langle \hat{\eta}(t) \hat{\eta}^\dagger(t') \rangle & = (n_\mathrm{th} + 1) \delta(t-t') , \\
\label{eq:etaProp2} \langle \hat{\eta}^\dagger(t) \hat{\eta}(t') \rangle & = n_\mathrm{th}  \delta(t-t') , 
\end{align}
and $\langle \hat{\eta}(t) \hat{\eta}(t') \rangle = 0$. 

Finally, to arrive at \eqref{eq:bEquation}, we have chosen the mechanical detuning $\Delta_m$ to match a shift in the mechanical resonance frequency due to the optomechanical interaction, {\it i.e.}, the optical spring effect. This choice can be viewed as simply the definition of $\tilde{\omega}_m$. Equivalently, it can be seen as a renormalization of the sideband frequency splitting $2 \delta$, which we in any case choose freely. This reflects the fact that the precise value of the mechanical frequency $\omega_m$ is not important in the setup we propose. 

Solving Equation \eqref{eq:bEquation}, using the noise properties \eqref{eq:zetaProp}, \eqref{eq:etaProp}, \eqref{eq:etaProp2}, and ignoring corrections of order $\tilde{\gamma}/\omega_m$, we find that Equations \eqref{eq:bCorrDef} and \eqref{eq:bCorrDef2} are valid, with the average phonon occupation number
\begin{align}
\label{eq:nm}
n_m & =  \frac{n_\therm + C_b + s C_r }{1 + (1 - s ) (C_r - C_b)} .
\end{align}
The two last terms in the numerator of Equation \eqref{eq:nm} represent heating due to Raman scattering of photons from the two drive frequencies to their lower sidebands, or equivalently, from radiation pressure shot noise. We also find that the off-diagonal mechanical correlation functions vanish in the limit $\tilde{\gamma}/\delta \rightarrow 0$ - see Appendix \ref{app:FiniteOverlap} for further details.
 
We now consider the limit of predominantly optical damping of the mechanical mode, {\it i.e.}, $C_r - C_b \gg 1$, which means $\beta < 1$, and the resolved-sideband limit $\omega_m \gg \kappa$, giving
\begin{align}
\label{eq:nmLimits}
n_m = \frac{n_m^{(0)} + \beta}{1 - \beta}
\end{align}
with
\begin{align}
\label{eq:nm0}
n_m^{(0)} & =  \frac{n_\therm}{C_r} + \left(\frac{\kappa}{4\omega_m} \right)^2 .
\end{align}
The parameter $n_m^{(0)}$ is the average phonon occupation number one would have for only red-detuned driving, {\it i.e.}, if $\beta = 0$, just as in standard optomechanical sideband cooling \cite{Wilson-Rae2007PRL,Marquardt2007PRL}. For sufficiently large cooperativity $C_r$, the first term can be made arbitrarily small. The second term in Equation \eqref{eq:nm0} is the usual limitation given by radiation pressure shot noise. We note that sideband cooling of modes of macroscopic mechanical systems have reached values of $n_m^{(0)}$ well below unity in a variety of experimental platforms, {\it e.g.}, superconducting circuits \cite{Teufel2011Nature}, suspended photonic crystals \cite{Chan2011Nature}, and dielectric membranes \cite{Purdy2015PRA,Underwood2015PRA}.

\begin{figure}[htb]
\includegraphics[width=.99\columnwidth]{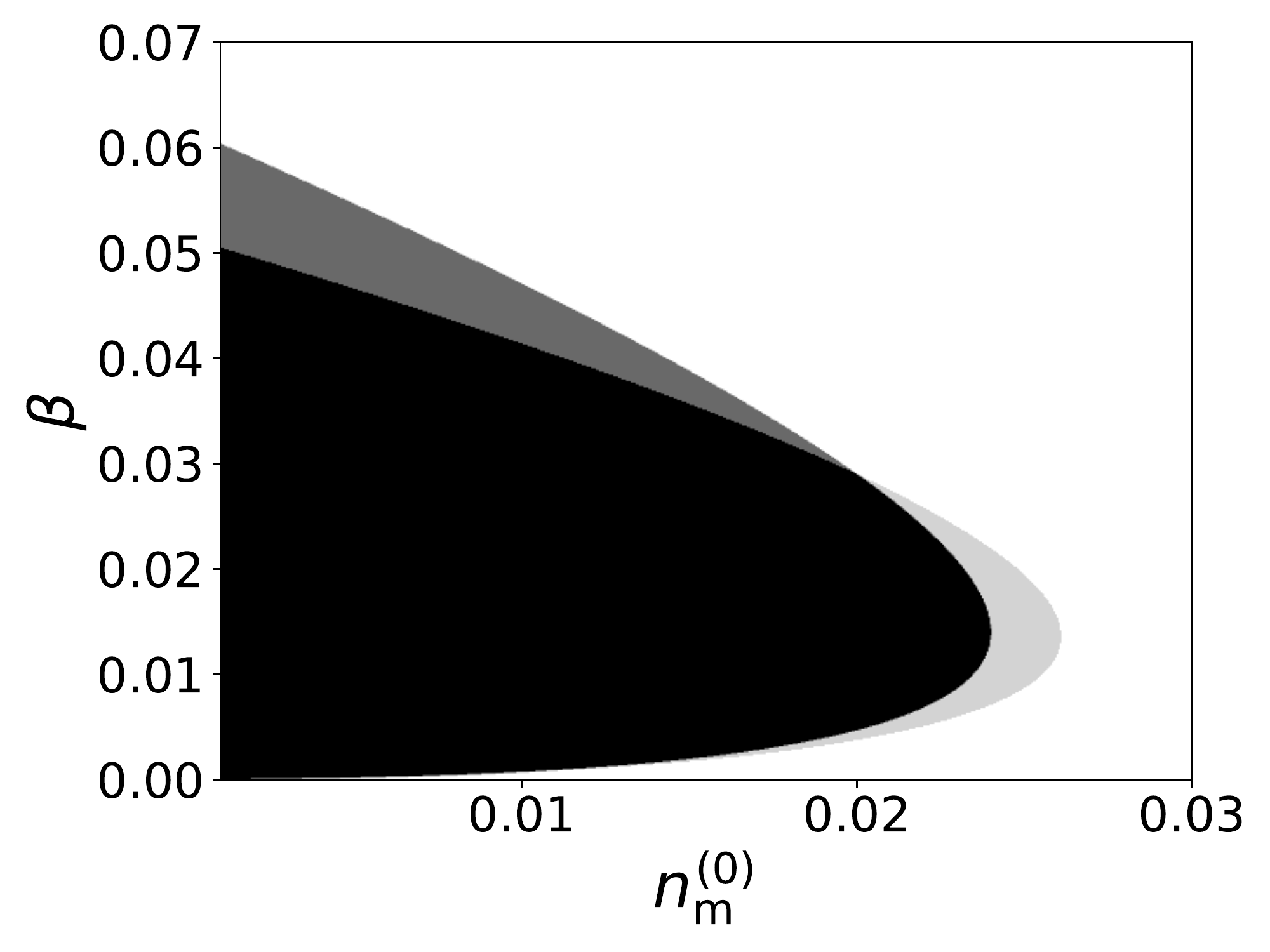}
\caption{{\it Dark gray}: Region of parameter space where the inequality \eqref{eq:KDelayIneq} is violated. {\it Light gray}: Region of parameter space where the inequality \eqref{eq:K0Ineq} is violated. {\it Black}: Region of parameter space where both inequalities are violated.}
\label{fig:KDelaynm0Plot}
\end{figure}
\begin{figure}[htb]
\includegraphics[width=.99\columnwidth]{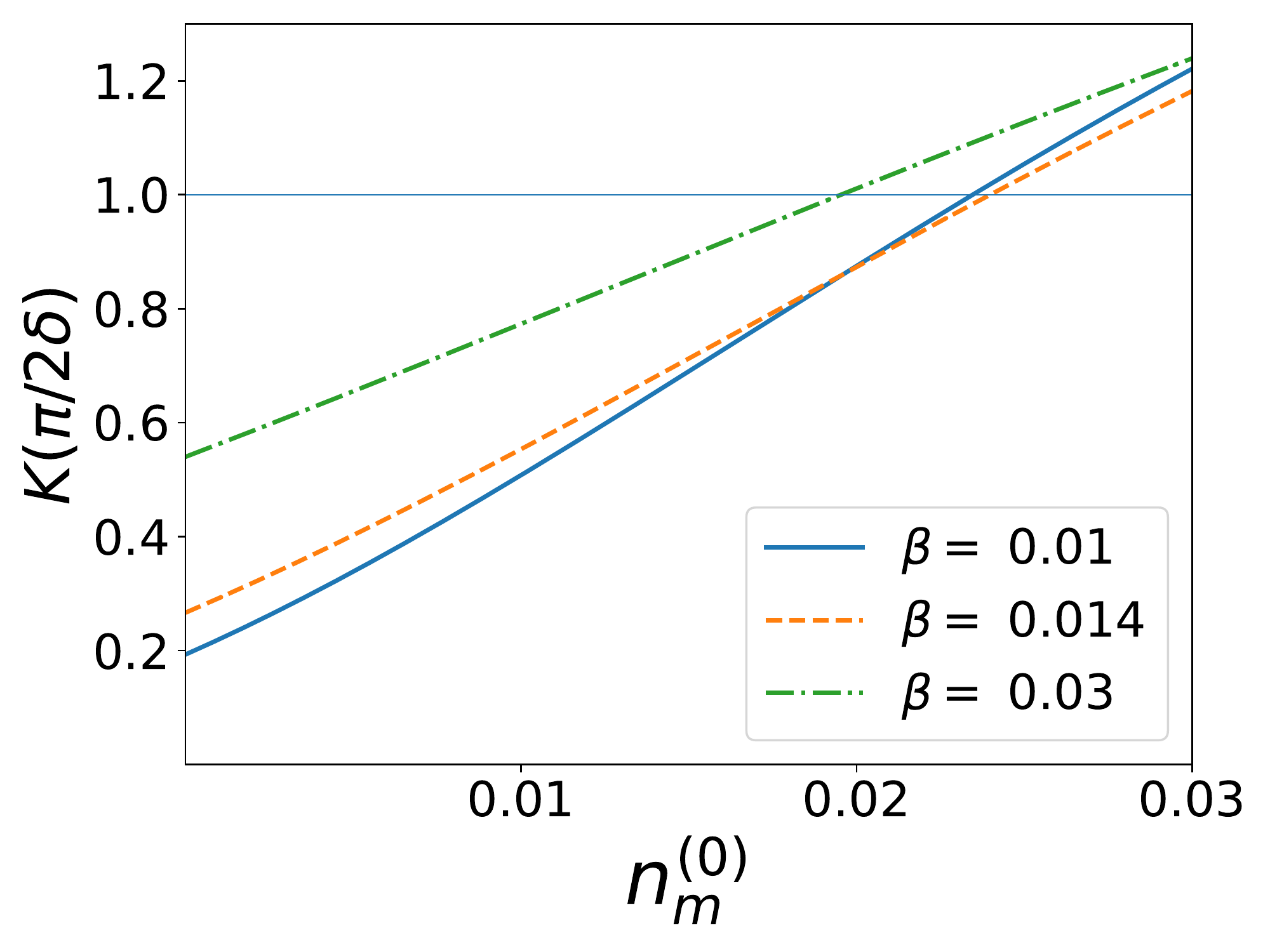}
\caption{The quantity $K(\pi/2\delta)$ as a function of $n_m^{(0)}$, i.e., the average phonon occupation number if $G_b$ were zero. The optimal choice of drive strength ratio for observing violation of \eqref{eq:KDelayIneq} is $\beta = 0.014$.}
\label{fig:Kvsnm0Plot}
\end{figure}
In Figure \ref{fig:KDelaynm0Plot}, we again plot the regions where the inequalities \eqref{eq:K0Ineq} and \eqref{eq:KDelayIneq} are violated, but now with $\beta$ and $n_m^{(0)}$ (not $n_m$) as the free parameters. The black region is the parameter regime where both inequalities are violated, the light gray region is where only the equal-time inequality \eqref{eq:K0Ineq} is violated, and the dark gray region is where only \eqref{eq:KDelayIneq} is violated. We also plot $K(\pi/2\delta)$ as a function of $n_m^{(0)}$ in Figure \ref{fig:Kvsnm0Plot}, which shows that a violation of \eqref{eq:KDelayIneq} can be observable for a sufficiently strong red-detuned drive and a system sufficiently far in the resolved sideband regime.

\section{Conclusion}
\label{sec:Conclusion} 

We have identified genuinely quantum features in the sideband photon statistics of an optomechanical cavity that is continuously driven. Compared to the standard optomechanical system with frequency filtered cavity output, the proposed setup is accessible simply by adding a second drive tone. Therefore, our results should be relevant to a variety of different experimental platforms.

We note that to violate the model-independent classical inequalities we have studied requires cooling of the mechanical mode to quite low occupation numbers, namely $n_m \lesssim 0.12$ when cooled by other means or $n_m^{(0)} \lesssim 0.02$ when relying on cooling intrinsic to the setup. However, the results presented can be useful for observing agreement with quantum theory also for higher occupation numbers, as long as one can verify the accuracy of the model by additional checks.

\acknowledgements{
We thank Yogesh Patil, Lucy Yu, Yiqi Wang, and Leon Loveridge for useful comments. KB acknowledges financial support from the Research Council of Norway (Grant No.~285616) through participation in the QuantERA ERA-NET Cofund in Quantum Technologies (project QuaSeRT) implemented within the European Union's Horizon 2020 Programme.
}

\appendix

\section{Details on filtering}
\label{app:Filtering}

\begin{figure}[htb]
\includegraphics[width=.99\columnwidth]{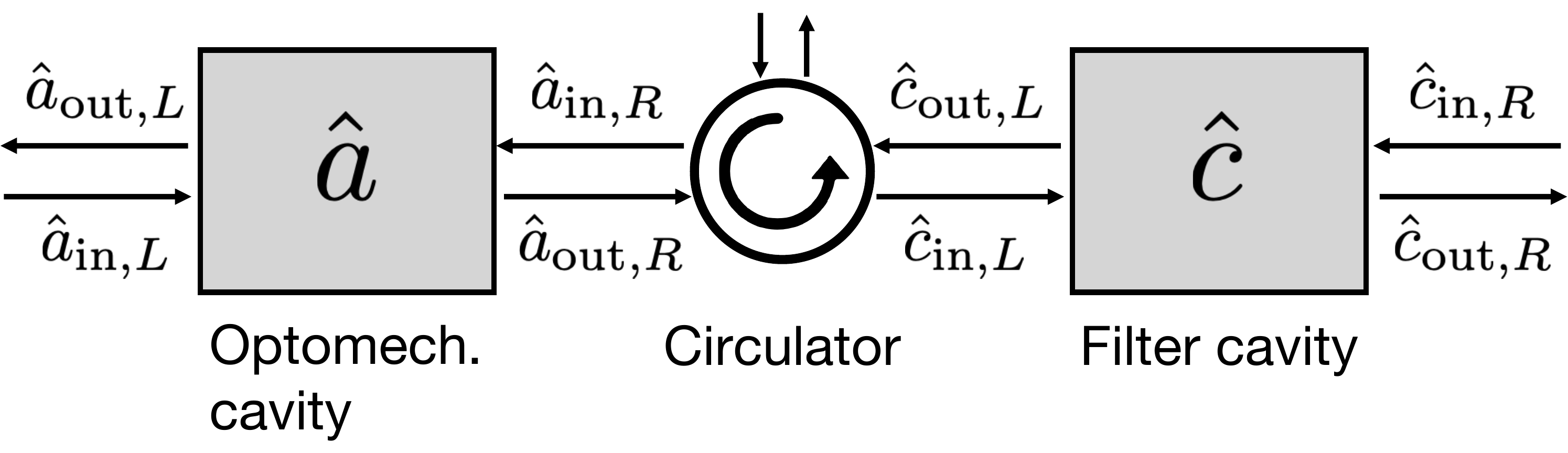}
\caption{Schematic overview of the relationship between the optomechanical cavity and the filter cavity/cavities.}
\label{fig:FilterDetails}
\end{figure}

To model the effect of the frequency filtering, we imagine a single filter cavity mode with photon annihilation operator $\hat{c}$ as shown schematically in Figure \ref{fig:FilterDetails}. We assume that the modes $\hat{a}$ and $\hat{c}$ have equal resonance frequencies. The right hand side output field from the optomechanical cavity mode $\hat{a}$ is 
\begin{align}
\label{eq:aoutR}
\hat{a}_{\mathrm{out},R}(t) = \sqrt{\kappa_R} \, \hat{a}(t) - \hat{a}_{\mathrm{in},R}(t)
\end{align}
where $\kappa_R$ is the contribution to the cavity linewidth coming from the decay through the mirror on the right. The circulator ensures that $\hat{a}_{\mathrm{in},R}$ is independent of the output $\hat{c}_{\mathrm{out},L}$ from the filter cavity and thus only represents vacuum noise. 

We also assume that the left hand side input field to the filter cavity is $\hat{c}_{\mathrm{in},L}(t) = \hat{a}_{\mathrm{out},R}(t)$, {\it i.e.}, that the circulator realizes a cascaded quantum system \cite{Gardiner1993PRL,Carmichael1993PRL}. We ignore any time delay due to the finite speed of light here, but this is not essential. The input field on the right hand side of the filter cavity, $\hat{c}_{\mathrm{in},R}$, is simply vacuum noise.

We denote the filter cavity decay rate $B$. In the Fourier representation, standard input-output theory for the empty filter cavity thus gives the right hand side output field 
\begin{align}
\label{eq:coutR}
\hat{c}_{\mathrm{out},R}[\omega] & =   \sqrt{B_L B_R} \chi_f(\omega) \left(\sqrt{\kappa_R} \, \hat{a}[\omega] - \hat{a}_{\mathrm{in},R}[\omega]\right) \notag \\
& + \left(B_R \chi_f(\omega)  - 1\right) \hat{c}_{\mathrm{in},R}[\omega] , 
\end{align}
where $B_L$ ($B_R$) is the contribution to the filter cavity decay rate from its left (right) mirror and 
\begin{align}
\label{eq:chif}
 \chi_f(\omega) & = \frac{1}{B/2 - i (\omega + \Delta_c)} .
\end{align}
If we now assume $\delta \ll B \ll \kappa, \omega_m$, we approximately find
\begin{align}
\label{eq:coutRApprox}
& \hat{c}_{\mathrm{out},R}[\omega] =  \frac{2 \sqrt{B_L B_R \kappa_R} }{B}   \left(\frac{B}{2} \chi_f(\omega) \hat{\zeta}[\omega] + \hat{a}_i[\omega] \right)  \notag \\
& + \left(B_R \chi_f(\omega)  - 1\right) \hat{c}_{\mathrm{in},R}[\omega] -  \sqrt{B_L B_R} \chi_f(\omega)  \hat{a}_{\mathrm{in},R}[\omega]  . 
\end{align}
One should now note that the vacuum noise $\hat{c}_{\mathrm{in},R}$ is uncorrelated with all other terms, and therefore cannot contribute to any normal ordered correlation function involving the output field $\hat{c}_{\mathrm{out},R}(t)$. In addition, although $\hat{a}_{\mathrm{in},R}$ can be correlated with the optomechanical cavity mode operator (see Appendix \ref{app:CalcDetails}), the {\it explicit} dependence on $\hat{a}_{\mathrm{in},R}$ in Equation \eqref{eq:coutRApprox} cannot contribute to a time ordered correlation function involving the output field $\hat{c}_{\mathrm{out},R}(t)$ since $\langle \hat{a}_{\mathrm{in},R}(t + \tau) \hat{a}(t) \rangle$ is nonzero only for $\tau < 0$ due to causality. 
 
Thus, when defining the filtered cavity vacuum noise $\hat{\zeta}_f(t)$ through its Fourier transform
\begin{align}
\label{eq:zetafDef}
\hat{\zeta}_f[\omega] & =  \frac{B}{2} \chi_f(\omega) \hat{\zeta}[\omega] , 
\end{align}
it is clear from Equation \eqref{eq:coutRApprox} that the photon statistics of the right hand side output field $\hat{c}_{\mathrm{out},R}$ is the same as that calculated by the operator $\hat{a}_f$ defined in Equation \eqref{eq:afDef}. We also point out that, since $ \hat{c}_{\mathrm{in},R}$ is vacuum noise, the measured photon statistics is the same as the photon statistics of the filter cavity mode $\hat{c}$, such that the Glauber-Sudarshan function $P(\alpha)$ referred to in the text can be thought of as a representation of the state of mode $\hat{c}$.

\section{Corrections to ideal limits}
\label{app:Corrections}

\subsection{Definitions}

We start by defining the normalized cavity response 
\begin{align}
\label{eq:tDef}
t(\omega) = \left(\frac{\kappa}{2}\right)^2 \left|\chi_c(\omega)\right|^2 ,
\end{align}
which measures how easy it is to put a photon in the cavity mode at a particular frequency. We also define the {\it effective} cooperativities 
\begin{align}
\label{eq:CtildeDef}
\tilde{C}_r = t(\delta) C_r \ , \ \tilde{C}_b = t(-\delta) C_b  ,   
\end{align}
which adjusts for the fact that the sidebands are not necessarily exactly at the cavity resonance frequency, as well as their ratio
\begin{align}
\label{eq:betatildeDef}
\tilde{\beta} = \frac{\tilde{C}_b}{\tilde{C}_r} = \frac{t(-\delta) }{t(\delta)} \beta . 
\end{align}

\subsection{Mechanical linewidth and average phonon number}
\label{app:Mech}

For nonzero $\delta/\kappa, |\Delta_c|/\kappa$, the effective mechanical linewidth becomes
\begin{align}
\tilde{\gamma} & = \gamma \left(1 + \left[t(\delta) -   t(-\Omega) \right] C_r  -  \left[ t(-\delta) -   t(\Omega) \right] C_b \right)
\end{align}
whereas the average phonon occupation number is corrected to
\begin{align}
\label{eq:nmCorr}
n_m & =  \frac{n_\therm + t(-\delta) C_b + t(-\Omega) C_r }{1 + \left[t(\delta) -   t(-\Omega) \right] C_r  -  \left[ t(-\delta) -   t(\Omega) \right] C_b} .
\end{align}
In the limit $\tilde{C}_r - \tilde{C}_b \gg 1$ and the resolved-sideband limit $\omega_m \gg \kappa$, we then get
\begin{align}
\label{eq:nmLimitsCorr}
n_m = \frac{n_m^{(0)} + \tilde{\beta}}{1 - \tilde{\beta}}
\end{align}
with
\begin{align}
\label{eq:nm0Corr}
n_m^{(0)} & =  \frac{n_\therm}{\tilde{C}_r} + \frac{t(-\Omega)}{t(\delta)} .
\end{align}

Let us briefly justify why we could neglect the term $\propto \mu$ in Equation \eqref{eq:bEquation}. This represents off-resonant two-phonon driving induced by the two drive tones separated by $2 (\tilde{\omega}_m - \delta)$. To second order in $\delta/\kappa, |\Delta_c|/\kappa$, we find $|\mathrm{Re} \, \mu| \ll |\mathrm{Im} \, \mu|$ and
\begin{align}
\label{eq:mu}
 \mathrm{Im} \, \mu & = \frac{2 \gamma \tilde{C}_r \sqrt{\tilde{\beta}}  \Delta_c }{\kappa} .
\end{align}
Second order perturbation theory in $\mu$ would give resonant corrections proportional to $|\mu|^2/\delta$, which should be compared to the effective linewidth:
\begin{align}
\label{eq:Corrmu}
\frac{|\mu|^2}{\tilde{\gamma} \delta} & \approx \left(\frac{4  G_b}{\kappa}\right)^2 \cdot \frac{\Delta_c^2}{\kappa \delta} \cdot \frac{1}{1/C_r + (1-s)(1-\beta)}    .
\end{align}
In the weak-coupling limit $C_r , C_b \ll 1$, this is then clearly negligible as long as $\Delta_c/\delta \lesssim 1$. Conversely, in the limit $C_r \gg 1$, the corrections are negligible as long as $\beta$ is not too close to 1, {\it i.e.}, as long as there is some effective sideband cooling. However, as we have seen, the ideal choice for observing the nonclassical features discussed is indeed the limit $\beta \ll 1$. Finally, we note that even in the case $\beta = 1$, neglecting $\mu$ is still justified as long as $C_r C_b \ll \kappa^2 \delta/(\gamma  \Delta_c^2)$.

\subsection{Finite sideband overlap}
\label{app:FiniteOverlap}

In the main text, we considered the limit $\tilde{\gamma}/\delta \rightarrow 0$, {\it i.e.}, stricly separated sidebands. In practice, we neglected the off-diagonal mechanical correlation functions, which for finite $\tilde{\gamma}/\delta$  become
\begin{align}
\label{eq:bbCorrs}
\langle \hat{b}(t + \tau) \hat{b}(t) \rangle & = e^{2 i \delta t} e^{-\tilde{\gamma}\tau/2} \sigma_m \\
\label{eq:bbCorrs2}
\langle \hat{b}^\dagger(t + \tau) \hat{b}^\dagger(t) \rangle & = e^{-2 i \delta t} e^{-\tilde{\gamma}\tau/2} \sigma^\ast_m 
\end{align}
 with
\begin{align}
\label{eq:sigmam}
 \sigma_m  & = - \frac{\gamma \sqrt{\tilde{C}_r \tilde{C}_b}}{\tilde{\gamma} + 2i \delta} .
\end{align}

\subsection{Calculation of second and third order coherence}
\label{app:CalcDetails}

Due to the system dynamics being linear, Wick's theorem gives that the normalized second and third order coherences, defined in Equations \eqref{eq:g2Def} and \eqref{eq:g3Def}, can be expressed as
\begin{align}
\label{eq:g2Wicks}
& g^{(2)}(t,t+\tau)   \\
& = 1 + \frac{|\langle a_f^\dagger(t+\tau) a_f(t) \rangle |^2 + |\langle a_f(t+\tau) a_f(t) \rangle |^2}{\langle a_f^\dagger(t) a_f(t) \rangle \langle a_f^\dagger(t+\tau) a_f(t+\tau) \rangle} \notag 
\end{align}
and
\begin{align}
\label{eq:g3Wicks}
& g^{(3)}(t , t ,t+\tau) = 4 g^{(2)}(t,t+\tau) + g^{(2)}(t,t) - 4  \\
& \qquad + 4 \, \mathrm{Re} \,  \frac{\langle a_f^2(t) \rangle^\ast  \langle a_f^\dagger(t+\tau) a_f(t) \rangle \langle a_f(t+\tau) a_f(t) \rangle }{\langle a_f^\dagger(t) a_f(t) \rangle^2 \langle a_f^\dagger(t+\tau) a_f(t+\tau) \rangle} . \notag
\end{align}
We still consider the limit $\tilde{\gamma}/\kappa \rightarrow 0$, {\it i.e.}, the limit where the cavity adiabatically follows the mechanical mode, but we now include corrections to the limits $\tilde{\gamma}/\delta \rightarrow 0$, $\delta/\kappa \rightarrow 0$, and $|\Delta_c|/\kappa \rightarrow 0$. 

To evaluate \eqref{eq:g2Wicks} and \eqref{eq:g3Wicks}, we need the correlation function
\begin{align}
\label{eq:aCorrDef}
& \langle \hat{a}_f^\dagger(t+\tau) \hat{a}_f(t) \rangle  = \langle \hat{a}_i^\dagger(t+\tau) \hat{a}_i(t) \rangle = e^{-\tilde{\gamma}\tau/2} \\
& \times \left\{e^{i\delta \tau} G_r^2 |\chi_c(\delta)|^2  \left(n_m - \frac{\gamma \tilde{C}_b}{\tilde{\gamma} - 2 i \delta}  \right) \notag \right. \\
& \left. +  \, e^{-i\delta \tau} G_b^2 |\chi_c(-\delta)|^2 \left( n_m + 1 - \frac{\gamma \tilde{C}_r}{\tilde{\gamma} + 2 i \delta} \right) \right\} , \notag
\end{align}
as well as the off-diagonal correlation function
\begin{align}
\label{eq:aCorrDef2}
\langle \hat{a}_f(t+\tau) \hat{a}_f(t) \rangle & = \langle \hat{a}_i(t+\tau) \hat{a}_i(t) \rangle +  \langle \hat{\zeta}_f(t+\tau) \hat{a}_i(t) \rangle  , 
\end{align}
where the first term becomes
\begin{align}
\label{eq:aiai}
& \langle \hat{a}_i(t+\tau) \hat{a}_i(t) \rangle  = - e^{-\tilde{\gamma}\tau/2} G_r G_b \chi_c(\delta) \chi_c(-\delta) \\
& \times \left\{e^{i\delta \tau}  \left(n_m - \frac{\gamma \tilde{C}_b}{\tilde{\gamma} - 2 i \delta}  \right) \notag \right. \\
& \left. +  \, e^{-i\delta \tau}  \left( n_m + 1 - \frac{\gamma \tilde{C}_r}{\tilde{\gamma} + 2 i \delta} \right) \right\} . \notag
\end{align}
Compared to the results presented in the main text, \eqref{eq:aCorrDef} and \eqref{eq:aiai} include corrections of order $\gamma \tilde{C}_{r(b)}/\delta$ (due to terms proportional to \eqref{eq:bbCorrs} and \eqref{eq:bbCorrs2}). They also contain corrections of order $\delta^2/\kappa^2$, $\delta \Delta_c/\kappa^2$, and $\Delta_c^2/\kappa^2$, which we ignored in the main text when approximating $\chi_c(\pm \delta) \approx 2/\kappa$. The latter corrections simply leads to replacing $\beta$ by $\tilde{\beta}$ in Equations \eqref{eq:g2Sol} and \eqref{eq:g3DelayTime}.

In the main text, we also neglected the last term in \eqref{eq:aCorrDef2}, {\it i.e.}, correlations between the cavity vacuum noise and the mechanical mode. This term becomes
\begin{align}
\label{eq:QuantCorr}
\langle \hat{\zeta}_f(t+\tau) \hat{a}_i(t) \rangle   & = i e^{-(\kappa/2 - i \Delta_c)\tau}  G_r G_b \chi_c(\delta) \chi_c(-\delta)  \delta  \chi_c(0)
\end{align}
in the limit $\tilde{\gamma} \ll \kappa$. This result cannot be found by using the approximate Equation \eqref{eq:aInner}, since it involves an off-resonant phonon (as illustrated in Figure \ref{fig:VirtualPhonon}c), but must rather be calculated starting from the original Langevin equations. We note that the correlation function \eqref{eq:QuantCorr} scales as $\delta/\kappa$ compared to the first term in \eqref{eq:aCorrDef2}. Note also that it decays at a rate $\kappa/2$, since it represents processes where two photons are created simultaneously by a virtual phonon transition.

By using the above expressions, one can show that the corrections to $g^{(2)}(\tau)$ are in fact only of second order in the parameters $\delta/\kappa, |\Delta_c|/\kappa$ for an arbitrary delay time $\tau$. While the corrections to $g^{(3)}(\tau)$ can generally be of first order in $\delta/\kappa$, it can be shown that at the delay times $\tau = 0$ and $\tau = \pi/2\delta$ on which we have focused, the corrections are in fact only of second order in this small parameter.

\subsection{Fully overlapping sidebands}
\label{sec:delta0}

Let us briefly comment on the special case $\delta = 0$, where the correlation functions \eqref{eq:bbCorrs} and \eqref{eq:bbCorrs2}) cannot be neglected. In this case, we have
\begin{align}
\label{eq:delta0Exp}
n_m - \frac{\gamma \tilde{C}_b}{\tilde{\gamma} - 2 i \delta}  & \approx \frac{\gamma n_\therm}{\tilde{\gamma}} \\
n_m + 1 - \frac{\gamma \tilde{C}_r}{\tilde{\gamma} + 2 i \delta} & \approx \frac{\gamma (n_\therm + 1)}{\tilde{\gamma}} , 
\end{align}
where we have neglected the outermost sidebands, {\it i.e.}, made the rotating wave approximation, valid for $\omega_m/\kappa \gg 1$. The consequence of this is that the results in Equations \eqref{eq:g2Sol} and \eqref{eq:g3EqualTime} are valid also in this case, but with the effective phonon number $n_m$ replaced by the bath occupation number $n_\therm$. This means that, unlike for $\delta \neq 0$, the intrinsic sideband cooling would not be of help for observing violation of the inequality \eqref{eq:K0Ineq} in this case.


\end{document}